\documentclass[adp,a4paper,fleqn%
]{w-art}
\usepackage{times,cite,w-thm}

\usepackage[]{graphicx}
\usepackage[english]{babel}
\usepackage{amsmath}
\usepackage{amssymb}
\usepackage{amsfonts}
\begin{document}

\DOIsuffix{theDOIsuffix}

\Volume{1} \Month{01} \Year{2007}

\pagespan{1}{}

\Receiveddate{XXXX} \Reviseddate{XXXX} \Accepteddate{XXXX}
\Dateposted{XXXX}



\title[Tools to constrain $f(R)$-gravity]{PPN limit and cosmological gravitational waves as tools to constrain $f(R)$-gravity}

\author[M. De Laurentis]{Mariafelicia De Laurentis\inst{1}
  \footnote{Corresponding author\quad
  E-mail:~\textsf{felicia@na.infn.it}}}
\address[\inst{1}]{Dipartimento di Scienze Fisiche\ Universit\`a ``Federico II'' di Napoli and INFN Sez. di Napoli
        Compl. Univ. Monte S. Angelo Ed. N, via Cinthia I- 80126 Napoli (Italy)}
        
\author[S. Capozziello]{Salvatore Capozziello\inst{1}}

\author[S. Nojiri]{Shin'ichi Nojiri\inst{2}}
\address[\inst{2}]{ Department of Physics, Nagoya University, Nagoya 464-8602, (Japan)}

\author[S. D. Odintsov ]{Sergei Odintsov\inst{3}}
\address[\inst{3}]{Institucio Catalana de Recerca i Estudis Avancats (ICREA) and
Institut de Ciencies de l Espai (IEEC-CSIC), Campus UAB, Facultat
de Ciencies, Torre C5-Par-2a pl, E-08193 Bellaterra, Barcelona,
(Spain).}

\begin{abstract}
We  discuss the  PPN Solar-System
constraints and the GW stochastic background considering some
recently proposed $f(R)$ gravity models which satisfy both
cosmological
and stability conditions. Using the definition of
PPN-parameters $\gamma$ and $\beta$ in terms of $f(R)$-models
 and the definition of scalar GWs, we
compare and discuss if it is possible to search for parameter
ranges of  $f(R)$-models  working at Solar System and GW
stochastic background scale. 
\end{abstract}

\maketitle

\section{Field equations and viable $f(R)$-model}
Let us start from the following action (see \cite{SFV})
\begin{equation}
\mathcal{S}=\mathcal{S}_g + \mathcal{S}_m =\frac{1}{k^{2}}\int
d^{4}x\sqrt{-g}\left[R+f(R)+\mathcal{L}_{m}\right]\,,\label{eq:action}\end{equation}
where we have considered the gravitational and matter
contributions and $k^2\equiv 16\pi G$. The non-linear $f(R)$ term
has been put in evidence with respect to the standard
Hilbert-Einstein term $R$ and $\mathcal{L}_{m}$ is the
perfect-fluid matter Lagrangian. The field equations are
\begin{equation}
\frac{1}{2}g_{\mu\nu}F(R)-R_{\mu\nu}F'(R)-g_{\mu\nu}\square F'(R)+
\nabla_{\mu}\nabla_{\nu}F'(R)=-\frac{k^{2}}{2}T_{\mu\nu}^{(m)}.\label{eq:motion}\end{equation}

Here $F(R)=R+f(R)$ and $T_{\mu\nu}^{(m)}$ is the matter
energy\,-\,momentum tensor. 
Action (\ref{eq:action}) can  be recast in a  scalar-tensor form.
By using the conformal scale transformation $g_{\mu\nu}\rightarrow
e^{\sigma}g_{\mu\nu}$ with $\sigma=-\ln\left(1+f'(R)\right)$, the
action can be written in the Einstein frame as follows
\cite{SFNS}:

\begin{equation}
\mathcal{S}_{E}=\frac{1}{k^{2}}\int d^{4}x\sqrt{-g}\left(R-\frac{3}{2}g^{\rho\sigma}\partial_{\rho}\sigma
\partial_{\sigma}\sigma-V(\sigma)\right),\label{eq:actionE}\end{equation}

where

\begin{equation}
V(\sigma)=e^{\sigma}g\left(e^{-\sigma}\right)-e^{2\sigma}f\left(g\left(e^{-\sigma}\right)
\right)=\frac{R}{F'(R)}-\frac{F(R)}{F'(R)^{2}}.\label{eq:potential}\end{equation}

The form of  $g\left(e^{-\sigma}\right)$ is given by solving
$\sigma=-\ln\left(1+f'(R)\right)=\ln F'(R)$ as
$R=g\left(e^{-\sigma}\right)$. The transformation
$g_{\mu\nu}\rightarrow e^{\sigma}g_{\mu\nu}$ induces a coupling of
the scalar field $\sigma$ with  matter.

Let us  consider now a class of $f(R)$ models which do not contain
cosmological constant and  are explicitly designed to satisfy
cosmological and Solar-System constraints in given limits of the
parameter  space. In practice, we choose a class of functional
forms of $f(R)$ capable of matching, in principle, observational
data. Firstly, the
cosmological model  should reproduce the  CMBR constraints in the
high-redshift regime (which agree with the presence of an
effective cosmological constant). Secondly, it should give rise to
an accelerated expansion, at low redshift, according to the
$\Lambda$CDM model. Thirdly, there should be sufficient degrees of
freedom in the parameterization to encompass low redshift
phenomena (e.g. the large scale structure) according to the
observations. Finally, small deviations from GR
should be consistent with Solar System tests. All these
requirements suggest that we can assume the  limits
\begin{equation}
\lim_{R\rightarrow\infty}f(R)={\rm constant},\qquad\, \lim_{R\rightarrow0}f(R)=0,
\end{equation}
which are satisfied by a general class of broken power law models,
proposed in \cite{Hu}, which are

\begin{equation}
F(R)=R-\lambda
R_{c}\frac{\left(\frac{R}{R_{c}}\right)^{2n}}{\left(\frac{R}{R_{c}}\right)^{2n}+1}
\label{eq:HS1}\end{equation}
 where parameters  $\{n$, $\lambda$, $R_{c}\}$
are constants which should be determined by experimental bounds.
\section{Constraining $f(R)$-models by PPN parameters}
The above model can be constrained at Solar System level by
considering the PPN formalism. This approach is extremely
important in order to test gravitational theories and to compare
them with GR. As it is shown in \cite{ppn-tot1,mpla}, one can
derive the PPN-parameters $\gamma$ and $\beta$ in terms of a
generic analytic function $F(R)$  and its derivative
\begin{equation}
\gamma-1=-\frac{F''(R)^{2}}{F'(R)+2F''(R)^{2}}\,,\qquad\,
\beta-1=\frac{1}{4}\left[\frac{F'(R)\cdot
F''(R)}{2F'(R)+3F''(R)^{2}}\right]\frac{d\gamma}{dR}\,.\label{eq:PPN}\end{equation}
These quantities have to fulfill the constraints  coming from the
Solar System experimental tests  summarized in Table I. They are
the perihelion shift of Mercury, the Lunar Laser
Ranging, the upper limits coming from the Very
Long Baseline Interferometry (VLBI) and the
results obtained from the Cassini spacecraft mission in the delay
of the radio waves transmission near the Solar conjunction.

\begin{table}[htb]
\begin{center}
\begin{tabular}{|c|c|}
\hline Mercury perihelion Shift&
$\left|2\gamma-\beta-1\right|<3\times10^{-3}$\tabularnewline
\hline \hline Lunar Laser Ranging &
$4\beta-\gamma-3=(0.7\pm1)\times10^{-3}$\tabularnewline \hline
Very Long Baseline Interferometer&
$\left|\gamma-1\right|<4\times10^{-4}$\tabularnewline \hline
Cassini Spacecraft&
$\gamma-1=(2.1\pm2.3)\times10^{-5}$\tabularnewline \hline
\end{tabular}
\end{center}
\caption{Solar System experimental constraints on the PPN
parameters.} \label{1}
\end{table}
By integrating last equations (\ref{eq:PPN}), one
obtains $f(R)$ solutions depending on $\gamma$ and $\beta$
which has to be confronted with $\gamma_{exp}$ and $\beta_{exp}$
plug into such equations the model and the experimental
values of PPN parameters and then
we will obtain algebraic constraints for the phenomenological
parameters $n$  and  $\lambda$ . Determining
the value from the equation for de Sitter
solutions according to the stability conditions
$F'(R)>0$ and $F''(R)>0$. Finally we obtain a
good sets of parameters for the model \cite{SFNS}:
\begin{itemize}
\item $\frac{R}{R_c}=3.38$\; , $n=1$\; , $\lambda=2$
\item $\frac{R}{R_c}=\sqrt{3}$\; , $n=2$\; , $\lambda>\frac{8}{3\sqrt{3}}$
\end{itemize}
\section{Constraining $f(R)$-models by stochastic  backgrounds of gravitational waves}
Also the stochastic background of GWs can be
taken into account in order to constrain models. This approach
could reveal  very interesting because production of primordial
GWs could be a robust prediction for any model attempting to
describe the cosmological evolution  at primordial epochs.
The main characteristics of the gravitational backgrounds produced
by cosmological sources depend both on the emission properties of
each single source and on the source rate evolution with redshift.
It is therefore interesting to compare and contrast the probing
power of these classes of  $f(R)$-models at  hight, intermediate
and zero redshift \cite{tuning,BCDF}. To this purpose,  let us  take into account the primordial
physical process which gave rise to a characteristic spectrum
$\Omega_{sgw}$ for the early stochastic background of relic scalar
GWs by which we can recast the further degrees of freedom coming
from fourth-order gravity. This approach can greatly contribute to
constrain viable cosmological models. The stochastic background of scalar GWs can be described in terms of a scalar field $\Phi$ and characterized by a dimensionless
spectrum We can write the energy
density of scalar GWs  in terms of the closure energy density of
GWs per logarithmic frequency interval as (\cite{maggiore})
\begin{equation}
\Omega_{sgw}(f)=\frac{1}{\rho_{c}}\frac{d\rho_{sgw}}{d\ln
f}\, ,\qquad\,\,\rho_{c}\equiv\frac{3H_{0}^{2}}{8\pi G}\label{eq: spettrodensita'
critica}\end{equation} is the  critical energy density of the
Universe, $H_0$ the today observed Hubble expansion rate, and
$d\rho_{sgw}$ is the energy density of the gravitational radiation
scalar part  contained in the frequency range from $f$ to $f+df$.
We are considering now standard units.

The calculation for a simple inflationary model can be performed
assuming that the early Universe is described by an inflationary
de Sitter phase emerging in the radiation dominated era \cite{relix}. The conformal metric element is
\begin{equation}
ds^{2}=a^{2}(\eta)[-d\eta^{2}+d\overrightarrow{x}^{2}+
h_{\mu\nu}(\eta,\overrightarrow{x})dx^{\mu}dx^{\nu}], \label{eq:
metrica}\end{equation} and a  GW with  tensor and scalar modes in
the $z+$ direction   is given by \cite{CCD}
\begin{equation}
\tilde{h}_{\mu\nu}(t-z)=A^{+}(t-z)e_{\mu\nu}^{(+)}+A^{\times}(t-z)e_{\mu\nu}^{(\times)}+\Phi(t-z)e_{\mu\nu}^{(s)}\,.\label{eq:
perturbazione totale}\end{equation}
The pure scalar component is
then
\begin{equation}
h_{\mu\nu}=\Phi e_{\mu\nu}^{(s)}\,,\label{eq: perturbazione
scalare}\end{equation} where $e_{\mu\nu}^{(s)}$ is the
polarization tensor. At lower frequencies, the spectrum is given by
\begin{equation}
\Omega_{sgw}(f)\propto f^{-2}.\label{eq: spettro basse
frequenze}\end{equation}
 It is
interesting to calculate the  corresponding strain, where interferometers like VIRGO, LIGO and LISA reach a maximum
in sensitivity.  The well known
equation for the characteristic amplitude \cite{maggiore},
adapted to the scalar component of GWs, can be used. It is
\begin{equation}
\Phi_{c}(f)\simeq1.26\times
10^{-18}\left(\frac{1Hz}{f}\right)\sqrt{h_{100}^{2}\Omega_{sgw}(f)},\label{eq:
legame ampiezza-spettro}\end{equation} and then we obtain the
values in the Table \ref{2}.

\begin{table}[htb]
\begin{center}
\begin{tabular}{|c|c|}
\hline
$\Phi_{c}(100Hz)<2\times10^{-26}$&
LIGO\tabularnewline
\hline
$\Phi_{c}(100Hz)<2\times10^{-25}$&
VIRGO\tabularnewline
\hline
$\Phi_{c}(100Hz)<2\times10^{-21}$&
LISA\tabularnewline
\hline
\end{tabular}
\end{center}
\caption{Upper limits on the expected  amplitude for the GW scalar
component by ground-based-interferometers LIGO-VIRGO and space-interferometer LISA.} \label{2}
\end{table}
At this point, using the upper bounds in Table \ref{2},
calculated for the characteristic amplitude of GW scalar
component, let us test the  $f(R)$-gravity models, considered in
the previous sections,  to see whether they are compatible both
with the Solar System and GW stochastic background.

Before starting with the analysis, taking into account the above
discussion,  we have that the GW scalar component is
derived considering
 \begin{equation}
\Phi=-\frac{\delta\sigma}{\sigma_{0}}\,, \label{eq:PHI} \qquad
\sigma=-\ln(1+f'(R))=\ln
F'(R)\,,\qquad\delta\sigma=\frac{f''(R)}{1+f'(R)}\delta R\,.
\end{equation}
Then we obtain a
good sets of parameters for the model \cite{SFNS}:
\begin{itemize}
\item $\frac{R}{R_c}=3.38$\; , $n=1$\; , $\lambda=2$
\item $\frac{R}{R_c}=\sqrt{3}$\; , $n=2$\; , $\lambda>\frac{8}{3\sqrt{3}}$
\end{itemize}
such sets of parameters are same as
bounds coming from the PPN  and  some sets reproduce
quite well both the PPN upper limits and the
constraints on the scalar component amplitude
of GWs. The results indicate that self-consistent
models could be achieved comparing experimental
data at very different scales without extrapolating
results obtained only at a given scale \cite{SFNS}.
\section{Conclusions}
The interesting feature, and the main result of this paper, is
that such sets of parameters are not in conflict with bounds
coming from the cosmological stochastic background of GWs. In
particular, some sets of parameters  reproduce quite well both the
PPN upper limits and the constraints on the scalar component
amplitude of GWs.

Far to be definitive, these preliminary results indicate that
self-consistent models could be achieved comparing experimental
data at very different scales without extrapolating results
obtained only at a given scale.

\end{document}